%% file: HIN-25-010_temp.tex
\documentclass[11pt,twoside,a4paper,cmspaper,final,collab]{cms-tdr}
\def\svnVersion{aed2966-D}\def\svnDate{2026/06/01}\def\cmsCernNoTag{CERN-EP-2026-153}\def\cmsCernDate{\today}\def\cmsMessage{Submitted to Physics Letters B}
\begin{document}\cmsNoteHeader{HIN-25-010}

\newlength\cmsTabSkip\setlength{\cmsTabSkip}{1ex}
\newlength\cmsFigWidth
\ifthenelse{\boolean{cms@external}}{\setlength\cmsFigWidth{0.99\columnwidth}}{\setlength\cmsFigWidth{0.6\textwidth}}
\ifthenelse{\boolean{cms@external}}{\newcommand{\cmsLeft}{upper\xspace}}{\newcommand{\cmsLeft}{left\xspace}}
\ifthenelse{\boolean{cms@external}}{\newcommand{\cmsRight}{lower\xspace}}{\newcommand{\cmsRight}{right\xspace}}
\ifthenelse{\boolean{cms@external}}{\providecommand{\cmsTable}[1]{\resizebox{\columnwidth}{!}{#1}}}{\providecommand{\cmsTable}[1]{#1}}
\newcommand {\nch}      {\ensuremath{N_{\text{ch}}}\xspace}
\newcommand {\dnchdeta} {\ensuremath{\rd N_{\text{ch}}/\rd \eta}\xspace}
\newcommand {\ampt}     {\textsc{ampt}\xspace}
\newcommand {\epos}     {\textsc{epos}\xspace}
\newcommand {\eposlhc}  {\textsc{epos lhc}\xspace}
\newcommand {\starlight}{\textsc{STARlight}\xspace}
\newcommand {\dpmjet}   {\textsc{dpmjet-III}\xspace}
\newcommand {\trajectum}{\textsc{Trajectum}\xspace}
\newcommand {\dr}       {\ensuremath{\Delta r}\xspace}
\newcommand {\vz}       {\ensuremath{v_{z}}\xspace}
\newcommand {\ncoll}    {\ensuremath{N_{\text{coll}}}\xspace}
\newcommand {\npart}    {\ensuremath{N_{\text{part}}}\xspace}
\newcommand {\avenpart} {\ensuremath{\langle N_{\text{part}} \rangle}\xspace}
\newcommand{\PbPb}{\ensuremath{\mathrm{PbPb}}\xspace}
\newcommand{\pp}{\Pp{}\Pp}
\newcommand{\pPb}{\Pp{}\ensuremath{\mathrm{Pb}}\xspace}
\newcommand{\OO}{\ensuremath{\mathrm{OO}}\xspace}
\newcommand{\CuCu}{\ensuremath{\mathrm{CuCu}}\xspace}
\newcommand{\AuAu}{\ensuremath{\mathrm{AuAu}}\xspace}
\newcommand{\XeXe}{\ensuremath{\mathrm{XeXe}}\xspace}
\newcommand{\AandA}{\ensuremath{\mathrm{AA}}\xspace}
\newcommand{\pA}{\Pp{}\ensuremath{\mathrm{A}}\xspace}
\newcommand{\pAu}{\Pp{}\ensuremath{\mathrm{Au}}\xspace}
\newcommand{\dAu}{\ensuremath{\mathrm{dAu}}\xspace}
\cmsNoteHeader{HIN-25-010}
\title{Centrality dependence of charged-hadron pseudorapidity distributions in oxygen-oxygen collisions at \texorpdfstring{$\sqrtsNN = 5.36\TeV$}{sqrt(sNN)=5.36 TeV}}
\date{\today}

\abstract{
We report the first measurement of charged-hadron pseudorapidity ($\eta$) distributions in oxygen-oxygen (\OO) collisions at a nucleon-nucleon center-of-mass energy of $\sqrtsNN = 5.36$\TeV. The data were recorded by the CMS experiment at the LHC in 2025. Primary charged-hadron yields are measured in the range $\abs{\eta}<2.4$ as a function of centrality (the overlap of the two nuclei). The results are compared with previous measurements in lead-lead (\PbPb) and xenon-xenon collisions at similar energies, as well as predictions from several Monte Carlo event generators and a hydrodynamic model. The charged-hadron pseudorapidity density in the midrapidity region ($\abs{\eta}<0.5$) is $\langle \rd N_{\text{ch}}/\rd \eta \rangle = 41.8 \pm 1.1\syst$ integrated over centrality and $135.0 \pm 4.0\syst$ for the most central (\ie, largest nuclear overlap) events. The hydrodynamic model \textsc{Trajectum} provides the best overall description of the data, particularly in central collisions. The particle density at midrapidity divided by the number of nucleons participating in the interaction in central \OO collisions is consistent with that observed in central \PbPb collisions at similar collision energy. While the overall energy-scaling behavior observed in other nucleus-nucleus collisions is preserved, the data exhibit deviations from simple participant and system-size scaling, highlighting the role of collision geometry and finite-size effects in light ion collisions.
}
\hypersetup{
pdfauthor={CMS Collaboration},
pdftitle={Centrality dependence of charged-hadron pseudorapidity distributions in oxygen-oxygen collisions at sqrt(s[NN]) = 5.36 TeV},
pdfsubject={CMS},
pdfkeywords={CMS, hadrons, multiplicity, spectra}}

\maketitle

\section{Introduction}
Ultrarelativistic heavy ion collisions provide a unique opportunity to investigate strongly interacting matter under extreme temperatures and energy densities, where a novel state of matter governed by quantum chromodynamics (QCD)---the quark-gluon plasma (QGP)---is formed~\cite{Busza:2018rrf}.
One of the most striking properties of the QGP is its near-perfect fluidity, characterized by an exceptionally low shear viscosity to entropy density ratio, as inferred from comparisons between experimental data and hydrodynamic model calculations.
Bulk particle production observables, in particular charged-particle pseudorapidity ($\eta$) distributions \dnchdeta, play a central role in constraining the initial-state entropy deposition and the subsequent hydrodynamic response of the system~\cite{Romatschke:2017ejr}.
The dependence of \dnchdeta on the collision geometry and center-of-mass energy provides sensitivity to initial-state effects, such as nuclear shadowing and gluon saturation~\cite{Albacete:2014fwa}, while also probing the interplay between soft- and hard-scattering processes in particle production~\cite{Kharzeev:2000ph}.
These measurements further serve as key inputs for the development and validation of comprehensive theoretical models~\cite{dEnterria:2011twh} and are essential for analyses of hard probes, where the subtraction of the underlying event for isolated photons and jets relies on a precise understanding of inclusive particle production~\cite{CMS:2024krd,Apolinario:2022vzg}.
Recent studies also suggest that charged-particle multiplicity may be influenced by the nuclear deformation of the colliding ions~\cite{Sharma:2025jhs}.
Consequently, precise measurements of \dnchdeta are vital for constraining models of QGP formation and the dynamics of heavy ion collisions.

In 2025, the CMS experiment collected the first data set of oxygen-oxygen (\OO) collisions at the CERN Large Hadron Collider (LHC) at a nucleon-nucleon center-of-mass energy of $\sqrtsNN=5.36\TeV$.
With an average of only about 10 nucleons participating in an interaction, \OO collisions occupy an intermediate regime between proton-lead (\pPb) and lead-lead (\PbPb) systems in terms of size.
Despite being significantly smaller than \PbPb collision systems, \OO collisions are expected to produce conditions compatible with the formation of a hot partonic medium~\cite{Citron:2018lsq}.

In addition, as a doubly magic nucleus with closed proton and neutron shells, $^{16}$O has a well-constrained density profile, reducing theoretical uncertainties in initial-state modeling compared to heavier, potentially deformed nuclei~\cite{Zhang:2024vkh}.

First measurements of \OO collisions at the LHC~\cite{ALICEFlowOO,ATLAS:2025nnt,CMS:2025tga} have already revealed features commonly associated with collective behavior similar to those observed in larger collision systems.
These results suggest that \OO collisions may create a medium with properties consistent with those of a QGP-like state, while still offering a controlled environment with reduced initial-state uncertainties.
However, precise measurements of charged-particle multiplicity and its pseudorapidity dependence remain essential for quantitatively constraining initial-state entropy production and testing the scaling properties of particle production across collision systems.
The \OO measurements can be directly compared with earlier copper-copper~(\CuCu)~\cite{PHOBOS:2007hfc} and gold-gold~(\AuAu)~\cite{PHOBOS:2005zhy,BRAHMS:2001llo} results from the BNL RHIC, as well as xenon-xenon (\XeXe)~\cite{CMS:2019gzk,ALICE:2018cpu} at $\sqrtsNN=5.44\TeV$ and \PbPb at $2.76\TeV$~\cite{ALICE:2010mlf,CMS:2011aqh}, $5.02\TeV$~\cite{ALICE:2015juo}, and $5.36\TeV$~\cite{CMS:2024ykx,ALICE:2025cjn} from the LHC. 
These earlier studies reveal an approximate power-law scaling of charged-particle production with \sqrtsNN that is largely independent of the colliding nuclei in nucleus--nucleus (\AandA) collisions, while proton-proton (\pp) and proton-nucleus (\pA) systems follow power laws with different exponents. 
The \OO data therefore provide a stringent new test of this scaling behavior and of phenomenological models tuned to existing measurements~\cite{Deng:2010xg,Pierog:2013ria,Lokhtin:2005px}.

In this Letter, we present the first measurement of the pseudorapidity density, \dnchdeta, of primary charged hadrons with $\abs{\eta} < 2.4$ in \OO collisions at $\sqrtsNN = 5.36\TeV$, providing constraints on final-state particle production and, through comparisons with theoretical models, on entropy production.
Primary charged hadrons are defined as prompt charged hadrons and decay products of particles with a proper decay length $c\tau < 1\unit{cm}$, where $c$ is the speed of light in vacuum and $\tau$ is the proper lifetime of the particle. 
Contributions from prompt leptons, decay products of longer-lived particles, and secondary interactions are corrected for and excluded.
This definition is identical to that used in previous CMS analyses~\cite{CMS:2010wcx,CMS:2010tjh,Khachatryan:2010nk,CMS:2011aqh,Chatrchyan:2014qka,CMS:2015zrm,CMS:2017shj,CMS:2019gzk}.
The results are presented as functions of charged-hadron $\eta$ and collision centrality, which quantifies the geometrical overlap of the two incoming ions~\cite{Miller:2007ri,dEnterria:2020dwq}.
Finally, the \OO data are compared with predictions from several event generators, enabling discrimination between competing models of initial conditions, nuclear modifications, and final-state interactions in light ion systems.
Tabulated results are provided in the HEPData record for this analysis~\cite{hepdata}.

\section{The CMS detector}
The central feature of the CMS apparatus is a superconducting solenoid of 6\unit{m} internal diameter, providing a magnetic field of 3.8\unit{T}. 
Within the solenoid volume are a silicon pixel and strip tracker covering the range $\abs{\eta} < 2.5$, a lead tungstate crystal electromagnetic calorimeter, and a brass and scintillator hadron calorimeter, each composed of a barrel and two
endcap sections. 
Forward calorimeters (HF), made of steel and quartz-fibers and located on either side of the interaction point, extend the pseudorapidity coverage to $\abs{\eta} < 5.2$.
Muons are detected in gas-ionization chambers embedded in the steel flux-return yoke outside the solenoid. 
The beam pickup timing for experiments (BPTX) devices, located around the beam pipe at a distance of 175\unit{m} from the interaction point on either side, provide precise information on the timing of the incoming beams.
A detailed description of the CMS detector, together with a definition of the coordinate system used and the relevant kinematic variables, can be found in Refs~\cite{Chatrchyan:2008aa,CMS:2023gfb}.

Charged hadrons are reconstructed using the silicon pixel detector installed during the Phase-1 upgrade~\cite{CMS:2012sda}. 
It consists of four concentric cylindrical shells (layers) in the barrel region (BPIX) and three disks on each side of the interaction point in the forward region (FPIX).
The BPIX and FPIX consist of a total of 1184 and 672 modules, respectively, and provide excellent position resolution with their $100{\times}150\mum$ pixels. 
The layers of the BPIX are denoted in increasing order of their radial distance from the beam axis, \ie, the layer closest to the beam axis is referred to as layer 1, the next closest layer is referred to as layer 2, and so on, while the disks of the FPIX are referred to in increasing order of their longitudinal distance from the nominal interaction point.

\section{Event selection}
\label{sec:eventselection}
During data taking, minimum-bias \OO collision events were recorded with a trigger that required signal above noise level in at least one side of the HF calorimeters and coincident signals in both BPTX detectors, corresponding to the arrival of beam bunches from both directions in the same bunch crossing. 

The recorded data sample corresponds to an integrated luminosity of approximately 2\mubinv. 
The average number of interactions per bunch crossing was about 0.15, and contributions from pileup interactions are suppressed by rejecting events with more than one reconstructed interaction vertex.
Offline, hadronic interactions are selected by demanding at least two energy deposits above 4\GeV in each HF calorimeter and at least one reconstructed vertex within 15\unit{cm} of the detector center along the beam axis ($\abs{v_{z}} < 15\unit{cm}$).  
The interaction vertex is identified using the tracklet-based charged-particle reconstruction algorithm summarized in Section~\ref{sec:analysis}. 
These requirements enrich the hadronic event sample while suppressing contributions from beam backgrounds and electromagnetic (EM) interactions.
An examination of data collected during bunch crossings without nuclear collisions shows that these selections reject all residual backgrounds from beam conditions, beam halo, and cosmic-ray muons.
After all criteria are applied, approximately 2 million events remain for the analysis.
The inefficiency of these selections is corrected in bins of tracklet multiplicity using simulated events generated by \HIJING (version 1.383)~\cite{Wang:1991hta}, and the associated model dependence (discussed below) is included as a systematic uncertainty.
The residual contribution from diffractive and nonhadronic interactions passing the event selection is estimated using simulated events generated by \eposlhc (version v3400)~\cite{Werner:2005jf,Pierog:2013ria} and \textsc{starlight} (v2.2)~\cite{Klein:2016yzr} interfaced with \textsc{dpmjet} (v3.0-5)~\cite{Roesler:2000he}, respectively.
This subtraction is performed in bins of both transverse energy ($E_\mathrm{T}=E/\cosh{\eta}$, where $E$ is the measured energy) deposited in the HF calorimeters and tracklet multiplicity, and its contribution is also included in the total event selection efficiency uncertainty.
The contribution from beam contamination by other ions, such as helium nuclei produced by oxygen
transmutation during beam circulation, was found to be negligible~\cite{Das:2025rqb}. 

\begin{table}[htb!]
    \centering
    \topcaption{
    Centrality intervals and corresponding \avenpart values for 5.36\TeV \OO collisions, reported up to 70\% centrality, while the 0--100\% interval corresponds to the fully centrality-integrated event sample.
    }
    \label{tab:npart}
    \begin{tabular}{cc}
Centrality interval [\%] & \avenpart \\ 
\hline
0--5 & $26.3 \pm 0.6$ \\
5--10 & $24.0 \pm 0.6$ \\
10--15 & $22.0 \pm 0.4$ \\
15--20 & $20.1 \pm 0.4$ \\
20--25 & $18.1 \pm 0.4$ \\
25--30 & $16.1 \pm 0.5$ \\
30--35 & $14.1 \pm 0.5$ \\
35--40 & $12.3 \pm 0.6$ \\
40--45 & $10.7 \pm 0.6$ \\
45--50 & $9.2 \pm 0.6$ \\
50--55 & $7.9 \pm 0.5$ \\
55--60 & $6.8 \pm 0.5$ \\
60--65 & $5.9 \pm 0.4$ \\
65--70 & $5.1 \pm 0.3$ \\
\hline 
0--100 & $10.9\pm 0.4$ \\
    \end{tabular}
\end{table}

Centrality is used to characterize the geometry of heavy ion interactions and is strongly correlated with the collision impact parameter and the number of participating nucleons, $\npart$, as described in the Glauber model of nuclear collision geometry~\cite{dEnterria:2020dwq}.
In this study, centrality is determined by the scalar sum of transverse energy recorded in the HF calorimeters, as in earlier CMS analyses~\cite{Chatrchyan:2011sx,CMS:2011aqh}. 
The transverse-energy spectrum, corrected for the overall event selection efficiency and EM contamination, is divided into percentile bins representing equal fractions of the total inelastic hadronic cross section~\cite{CMS:2011aqh}.  
By convention, smaller percentiles correspond to more central (small impact parameter) collisions.
Because event-selection inefficiencies and electromagnetic contamination mainly affect the most peripheral collisions, the calibration is anchored at the $E_\mathrm{T}$ value corresponding to 80\% of the hadronic cross section, where the event selection efficiency is unity and the electromagnetic contamination is negligible. 
The position of this reference point is determined with a precision of about 10\%, and its uncertainty is propagated to the final results.
The corrected $E_\mathrm{T}$ distribution is fit using a two-component ansatz.
In this approach, particle production is assumed to arise from a combination of soft interactions, which scale with the number of participating nucleons ($\npart$), and hard-scattering processes, which scale with the number of binary nucleon–nucleon collisions ($\ncoll$). 
The contributions from $\npart$ and $\ncoll$ derived from the Monte Carlo (MC) Glauber model~\cite{Loizides:2025ule} are combined into an effective number of particle production sources, which is then convolved with a gamma distribution.
This parameterization accounts for event-by-event fluctuations in the transverse energy deposited per source, allowing the model to accurately reproduce the shape of the $E_\mathrm{T}$ spectrum over a wide range.
The fit also includes a constraint requiring the 80\% reference point of the $E_\mathrm{T}$ distribution in the model to be consistent, within uncertainties, with that measured in data.
The $\npart$ values are obtained in the same percentile intervals as the data. 
Following Ref.~\cite{Loizides:2025ule}, the Glauber calculation assumes a nucleon-nucleon inelastic cross section of $68.0 \pm 1.2\unit{mb}$ and $\omega=0.3\pm0.3$ for the parameterization of the nucleon-nucleon overlap.
The uncertainty in $\npart$ is estimated from the spread of available oxygen nuclear density profiles and from variations of the fit range between 60\% and 94\% of the $E_\mathrm{T}$ distribution.
The resulting $\npart$ values and their uncertainties are provided in Table~\ref{tab:npart}. 

\section{Analysis}
\label{sec:analysis}
The \dnchdeta measurement employs tracklets, each constructed from a pair of pixel clusters recorded in two distinct layers (or forward disks) of the silicon pixel detector.
A cluster is defined as a contiguous group of pixels whose combined charge exceeds the readout threshold.
When both clusters originate from the same charged particle, the differences in pseudorapidity ($\Delta\eta$) and azimuthal angle ($\Delta\phi$, in radians), computed with respect to the primary vertex, are small.
By requiring $\abs{\Delta\eta}$ and $\abs{\Delta\phi}$ to fall within predefined selection windows, the analysis efficiently selects tracklets that correspond to primary charged hadrons while suppressing combinatorial backgrounds.  
Details of the tracklet and vertex reconstruction algorithms can be found in Ref.~\cite{CMS:2011aqh}.

This analysis employs nine distinct tracklet categories, defined by the pixel layer pairs used to form each cluster pair.  
Six categories combine clusters from the four barrel layers in all possible combinations of two distinct barrel layers, while the remaining three are constructed from pairs of the three forward disks.  
Mixed barrel–disk combinations are excluded because they are more susceptible to ``looper'' background from low transverse momentum (\pt) charged particles that curl in the magnetic field and produce multiple clusters within a single layer.
After applying the corrections described below, the \dnchdeta distributions from all nine tracklet categories are averaged and then symmetrized about $\eta=0$ to obtain the final result.

Because each tracklet category has a different sensitivity to the underlying charged-hadron \pt spectrum, the individual \dnchdeta distributions also serve as independent cross-checks of systematic effects.
The lowest accessible charged-hadron \pt values are reached by tracklets formed from the innermost barrel layers, while the remaining categories provide complementary measurements with higher effective \pt thresholds.
During data taking, a small fraction of detector modules were found to exhibit a significantly different response in data compared to MC simulations.
These modules are identified and excluded from the analysis, and their effects are accounted for in the acceptance correction.

From the two clusters in a tracklet pair, a measure of angular distance 
\begin{equation}
\dr = \sqrt{{\left(\eta_{{i}} - \eta_{{j}}\right)^{2} + \left(\phi_{{i}} - \phi_{{j}}\right)^{2}}}
\end{equation}
is defined.  
Here, $\eta_{{i}({j})}$ is the pseudorapidity of the cluster in the ${i}({j})$-th layer or disk with respect to the primary vertex position, and $\phi_{{i}({j})}$ is defined similarly for the azimuthal angle. 
Tracklets caused by combinatorial background are reconstructed from clusters originating from different charged particles and are found to dominate at $\dr > 0.5$ in the simulation. 
To increase the purity of the tracklet sample and to reduce the sensitivity to the modeling of combinatorial background effects, only tracklets having $\dr < 0.5$ are used in the analysis. 
This selection also suppresses the looper background related to low-\pt charged particles. 
After the tracklet selection, corrections are applied for the background fraction (including misreconstructed tracklets and nonprimary charged particles), geometric acceptance, reconstruction efficiency, and event selection efficiency.
These corrections include an extrapolation to zero charged-hadron \pt to account for the finite minimum \pt of reconstructed tracklets.
The minimum charged-hadron \pt accessible with this method is 50\MeV, corresponding to tracklets formed using the two innermost barrel pixel layers.
The correction factors are derived from simulated samples of the \HIJING and \eposlhc generators interfaced with \GEANTfour~\cite{Agostinelli:2002hh} to model the detector response.

The $\dr$ spectrum from the \HIJING sample is found to best reproduce the data in the region where combinatorial background is significant, \ie, for $0.4<\dr<0.5$, and is therefore used for the nominal correction factors.
The simulated samples from other event generators are used to study systematic uncertainties. 
The correction factors are calculated as functions of $\eta$, \vz, and the tracklet multiplicity.

\section{Systematic uncertainties}
The uncertainties resulting from various systematic effects affecting the measurement are evaluated as follows:

\begin{description}

\item [\vz consistency.]
The potential mismodeling of the alignment of the pixel detector in the simulation affects the correction factors. 
This effect is studied by comparing the results obtained in different \vz intervals.
The resulting systematic uncertainty, which depends on centrality and $\eta$, ranges from 1\% to 4\%.

\item [Tracklet selection.]
The tracklet selection criteria affect the minimum \pt and signal-to-background ratio of reconstructed tracklets. 
The sensitivity of correction factors to selection criteria is examined by varying the nominal \dr requirement by $\pm 0.1$.
Since particles with lower \pt exhibit larger \dr values, this variation changes the effective minimum accessible \pt from about 50\MeV to about 100\MeV for tracklets formed from the first and fourth barrel pixel layers. 
Despite this relatively large variation for a single tracklet category, the impact on the final result is reduced by the small particle yield in this low-\pt region, the associated correction factors, and the averaging over nine tracklet categories with different intrinsic \pt sensitivities. 
The assigned uncertainty is less than 1\%.
\item [Deviation from averaged and symmetrized results.] 
In any given $\eta$ range, measurements can be made using multiple tracklet combinations. 
The maximum deviation of the measurements obtained with individual tracklet combinations from the final averaged and symmetrized result is assigned as a systematic uncertainty and varies with centrality and $\eta$ from 1\% to 9\%.

\item [Model dependence of the corrections.]
The model dependence of the correction factors for tracklet efficiency, acceptance, and event selection efficiency is evaluated using alternative sets of corrections derived from \eposlhc and \ampt (version 1.26t5)~\cite{Lin:2004en}, with and without string melting. 
These differences affect the correction for leptons and the extrapolation of the measured spectra to $\pt = 0$.
The maximum deviation from the nominal result is assigned as a systematic uncertainty and varies with centrality and $\eta$, ranging from 1\% to 7\%.

\item [HF energy thresholds.] 
The potential difference in the HF response between data and simulation, which affects the event selection efficiency correction, is studied by varying the selection thresholds.
At least two energy deposits above 4\GeV in each HF calorimeter are required in the nominal analysis. 
The analysis is performed by requiring three towers above 3\GeV and two towers above 5\GeV in each HF calorimeter, leading to systematic uncertainties of 0.5\%--1\% in the \dnchdeta results. 

\item [Single diffractive fraction.] 
The model dependence of the single diffractive fraction is studied by using alternative values derived from \PYTHIAviii~\textsc{Angantyr}~\cite{Bierlich:2018xfw}. 
The resulting differences in the observables are below 1\% and are assigned as a systematic uncertainty.

\item [Centrality calibration.] 
The determination of event centrality depends on the hadronic event selection efficiency and on contamination from EM processes. 
Since the inefficiency primarily affects the most peripheral collisions, its uncertainty results in a migration of events between centrality intervals. 
To evaluate the impact on the final results, alternative centrality calibrations are derived by varying the event selection efficiency within its uncertainty and are used to recategorize the data. 
The corresponding assigned systematic uncertainty increases from 0.5\% in central collisions to 8\% in peripheral events.
\end{description}
A summary of the systematic uncertainties affecting the measurement of charged-hadron multiplicities as functions of $\eta$ and centrality is given in Table~\ref{tab:tab1}. 
The individual contributions are summed in quadrature to obtain the total systematic uncertainty. 
The uncertainty associated with the variation among tracklet combinations is dominant, in particular at large $\abs{\eta}$, where acceptance and low-\pt\ effects are largest, while the centrality calibration uncertainty dominates in peripheral collisions.
The statistical uncertainties are below 0.1\% in all $\eta$ and centrality intervals and are therefore neglected throughout this Letter.

\begin{table}[htbp!]
  \centering
    \topcaption{Sources of systematic uncertainty affecting the measurement of
    charged-hadron multiplicities as functions of $\eta$ and centrality classes in \OO collisions at $\sqrtsNN = 5.36\TeV$. For each source, the quoted range reflects the spread of the uncertainty across individual $\eta$ bins and centrality intervals; no averaging over $\eta$ or centrality is implied.}
    \label{tab:tab1}
    \cmsTable{
    \begin{tabular}{lc}
      Source                                          & [\%]          \\
    \hline
      Consistency between tracklet combinations       & 1--9      \\
      Primary vertex position consistency             & 1--4      \\
      Tracklet selection                              & $<$1     \\
      Model dependence                                & 1--7      \\
      Event selection                                 & $<$1     \\
      Single diffractive fraction                     & $<$1     \\
      Centrality calibration                          & 0.5--8    \\\hline
      Total systematic uncertainty                     & 3--10     \\ 
    \end{tabular}
    }
\end{table}

\section{Results}
\begin{figure}[thb!]  
  \centering
    \includegraphics[width=\cmsFigWidth]{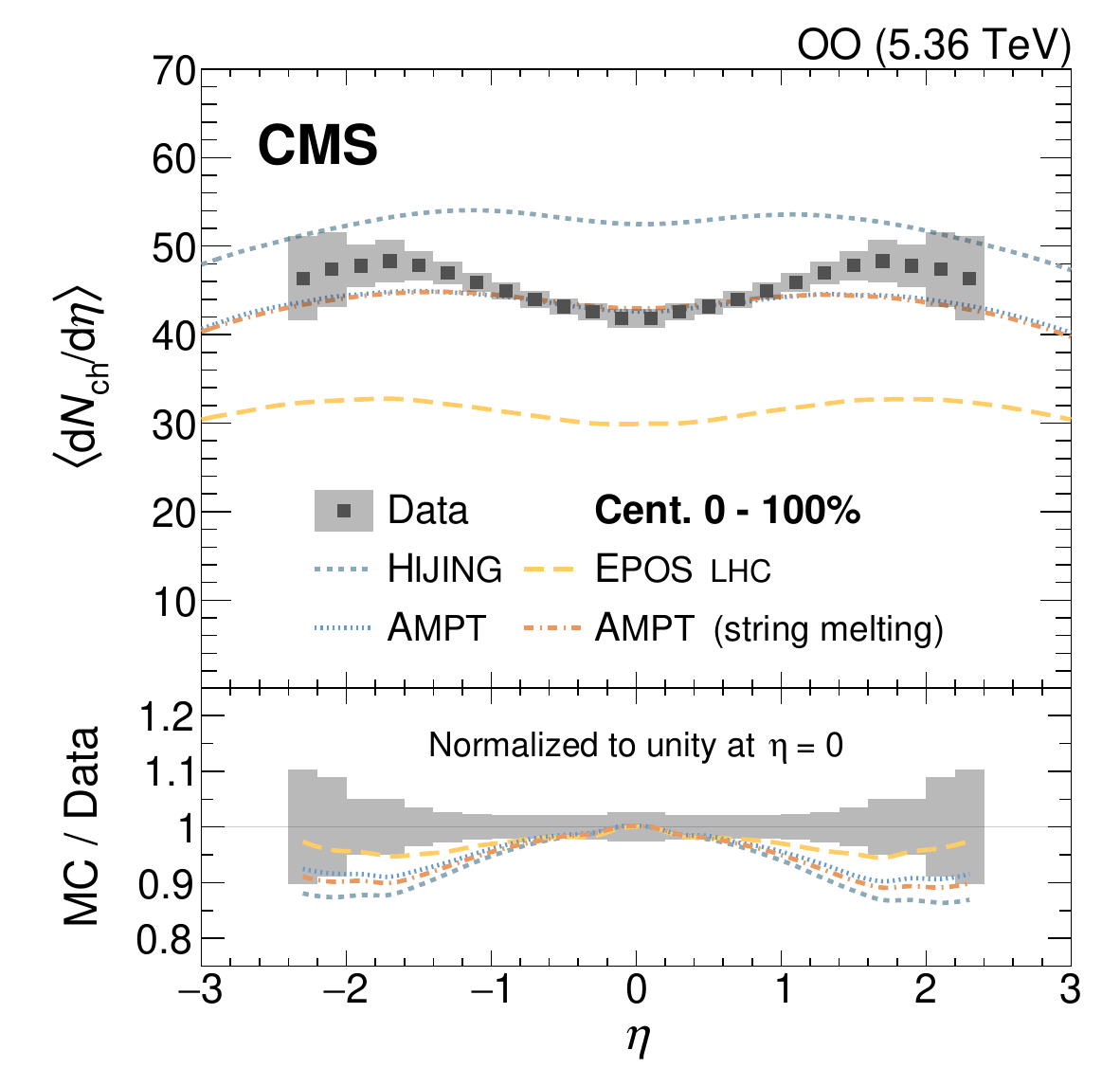}
    \caption{
    The \dnchdeta distribution in \OO collisions at $\sqrtsNN = 5.36\TeV$ for the 0--100\% centrality class. 
    The result is symmetrized about $\eta=0$. 
    Predictions from the \HIJING~\cite{Wang:1991hta}, \eposlhc~\cite{Werner:2005jf,Pierog:2013ria}, and \ampt~\cite{Lin:2004en} event generators are also shown. 
    The lower panel shows the ratios of the model predictions to the data, normalized to unity at midrapidity. 
    In the upper panel, the gray boxes indicate the total systematic uncertainties, while statistical uncertainties are negligible. 
    In the lower panel, the boxes represent the relative uncertainties of the data.}
    \label{fig:fig1}
\end{figure}

Figure~\ref{fig:fig1} shows the charged-hadron $\eta$ distribution for the 0--100\% centrality class.
The average charged-hadron pseudorapidity density in the midrapidity region ($\abs{\eta}<0.5$) is measured to be $\langle \rd N_{\text{ch}}/\rd \eta \rangle = 41.8 \pm 1.1\syst$.
The results are compared to predictions from the \HIJING, \eposlhc, and \ampt event generators. 
The \HIJING generator is based on perturbative QCD processes supplemented with a model for soft interactions and includes nuclear effects such as shadowing and jet quenching~\cite{Wang:1991hta}. 
The \eposlhc generator is a Gribov--Regge multiple-scattering framework that incorporates collective effects and hydrodynamic evolution, tuned to reproduce LHC data~\cite{Pierog:2013ria}. 
The \ampt generator combines the \HIJING event generator with Zhang's parton cascade procedure~\cite{Zhang:1997ej} and the ART model~\cite{Li:1995pra} for the final stage of hadronic interactions.
In the default \ampt, the excited matter is treated as strings that hadronize via the Lund string fragmentation process. 
In the string-melting version, however, these strings are converted into partons via an intermediate step in which the would-be hadrons are decomposed before coalescence into hadrons~\cite{Lin:2014tya}.
The two \ampt configurations give very similar predictions for the observables considered here.

None of the MC generators examined here are able to describe both the shape and magnitude of the charged-hadron distribution across the full range of $\eta$ probed.
At midrapidity, only \ampt agrees with the data, while in the forward region both \ampt and \HIJING are consistent with the data within uncertainties.
The predictions of \eposlhc significantly undershoot the data across all $\eta$ values.
The lower panel of Fig.~\ref{fig:fig1} shows the model-to-data ratios, normalized to unity at $\eta=0$ to highlight differences in the shape of the $\eta$ distributions independently of the overall normalization.
After accounting for normalization differences, \eposlhc provides the best description of the shape of the $\eta$ distribution among the considered generators. 
Similar conclusions were reached in studies of \XeXe collisions at $\sqrtsNN = 5.44\TeV$ by the CMS~\cite{CMS:2019gzk} and ALICE~\cite{ALICE:2018cpu} Collaborations.

\begin{figure}[hbt!]  
  \centering
    \includegraphics[width=\cmsFigWidth]{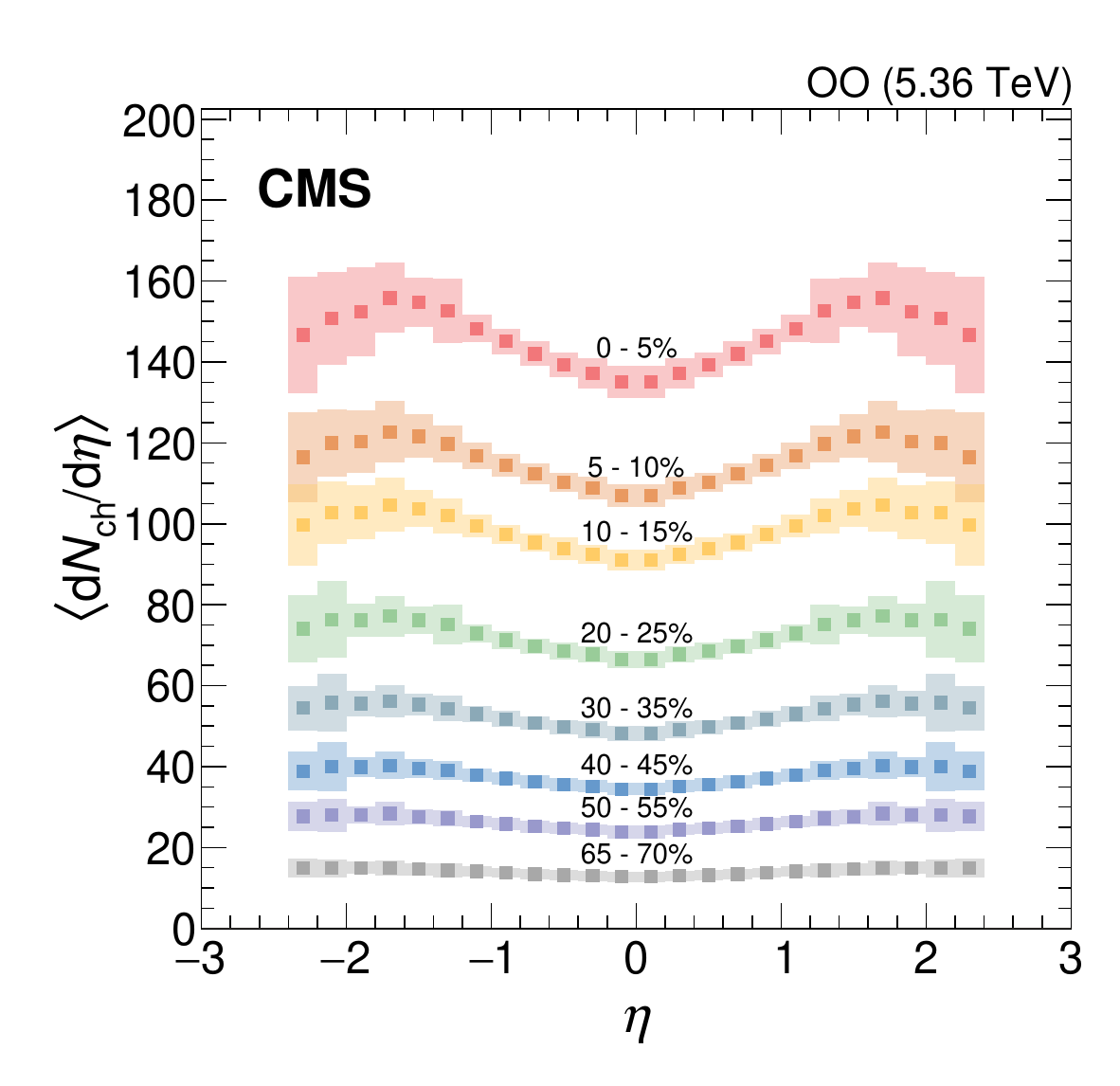}
    \caption{
    The \dnchdeta distribution in \OO collisions at $\sqrtsNN = 5.36\TeV$ for several centrality classes. 
    In all cases, the data are symmetrized about $\eta=0$. 
    Boxes indicate the total systematic uncertainties, while statistical uncertainties are negligible. 
    \label{fig:fignew}}
\end{figure}

To further explore the centrality dependence of particle production, the \dnchdeta distributions are measured in the finer centrality intervals listed in Table~\ref{tab:npart}.
Example distributions for a variety of centrality classes ranging from 0--5\% to 65--70\% are shown in Fig.~\ref{fig:fignew}. The observed $\dnchdeta$ values range from about 15 to 150.

\begin{figure}[hbt!]
  \centering
    \includegraphics[width=0.49\textwidth]{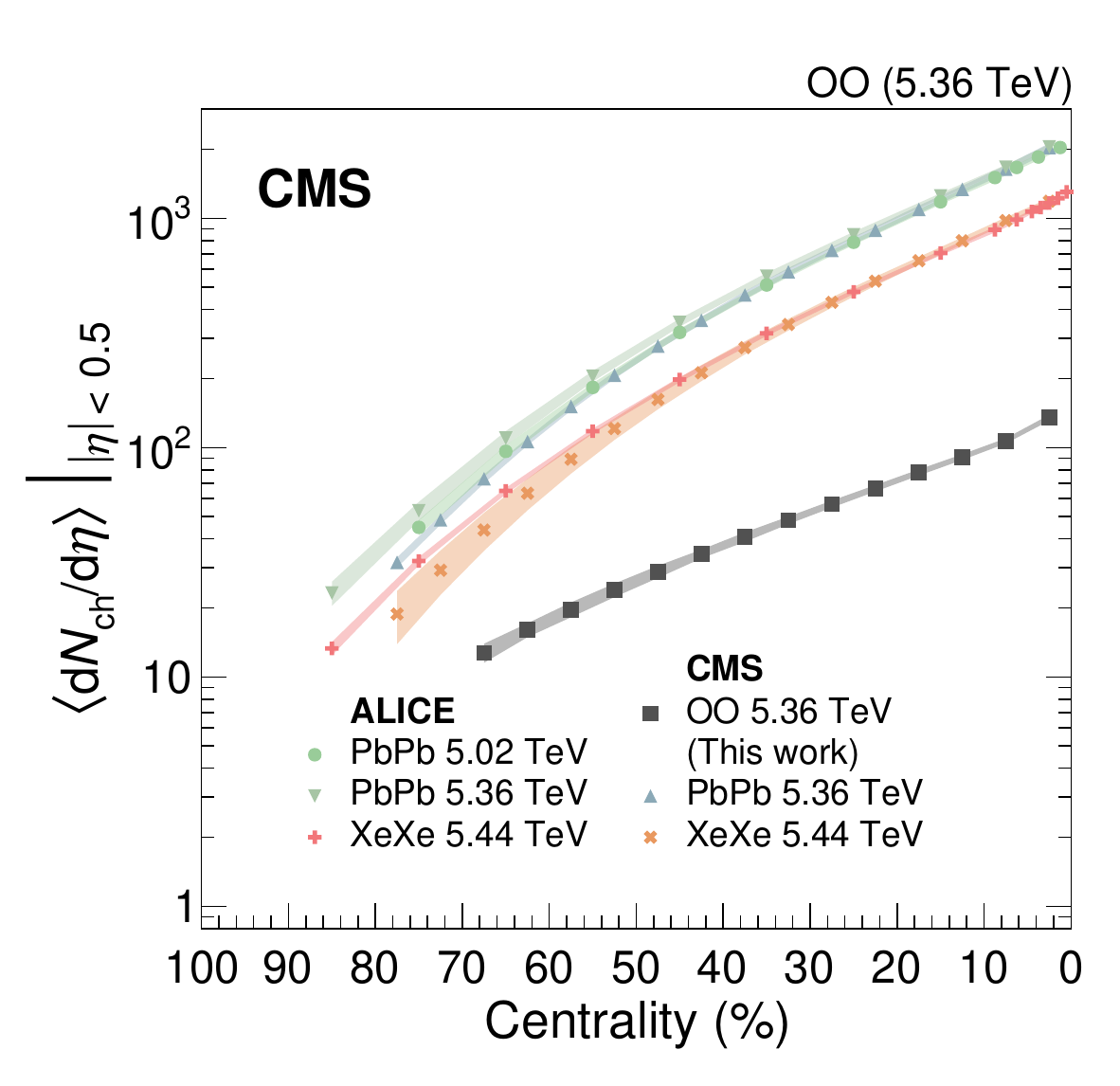}
    \includegraphics[width=0.49\textwidth]{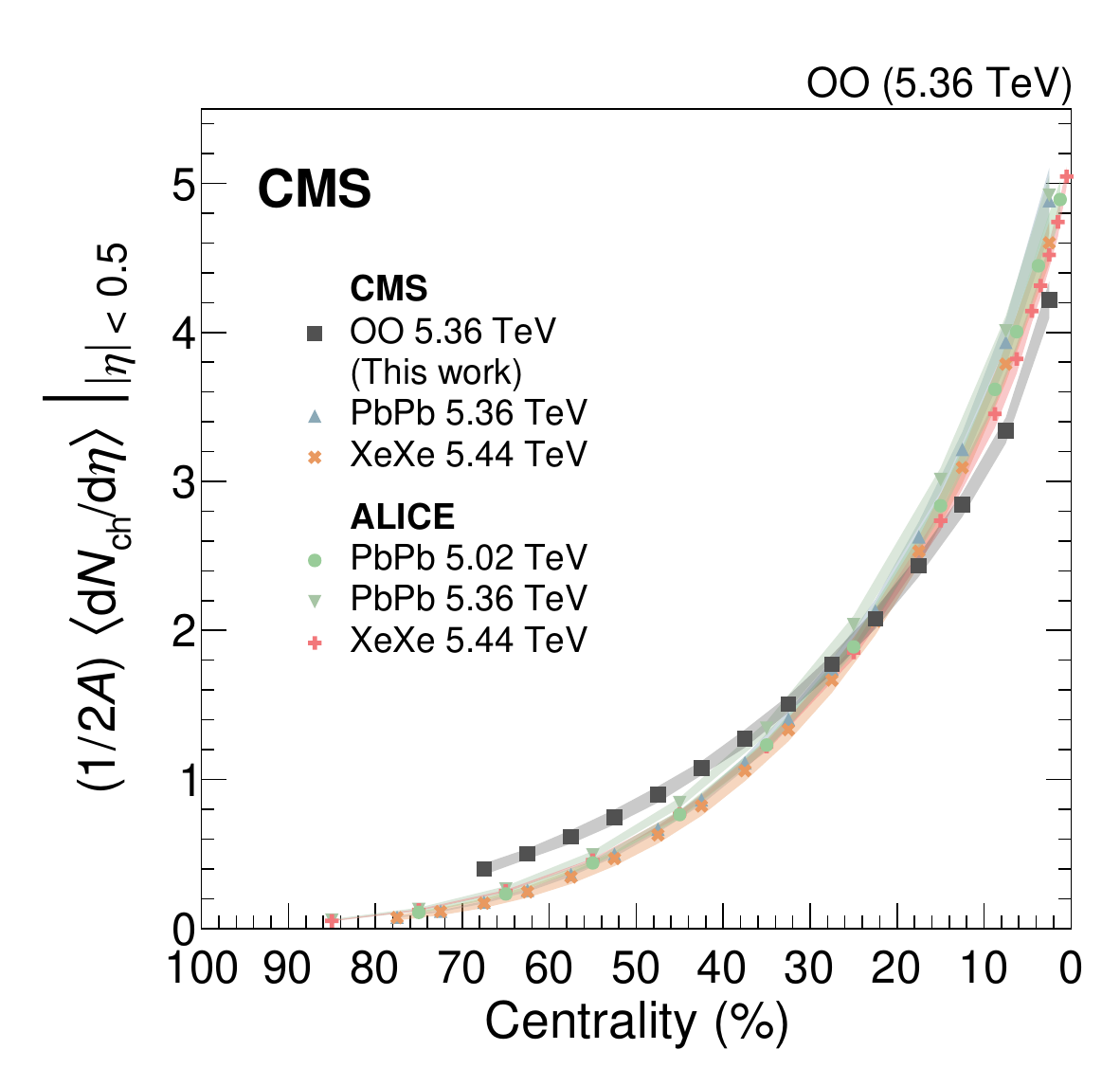}
    \caption{
    Charged-hadron \dnchdeta at midrapidity in \OO collisions at $\sqrtsNN = 5.36\TeV$ as a function of event centrality, shown for the measured values (\cmsLeft) and normalized by $2A$ (\cmsRight), where $A$ is the mass number of the colliding nuclei. 
    The results are compared with measurements in \PbPb and \XeXe collisions at similar collision energies by the CMS~\cite{CMS:2019gzk,CMS:2024ykx} and ALICE~\cite{ALICE:2015juo,ALICE:2018cpu,ALICE:2025cjn} Collaborations.
    The bands around the data points denote the total systematic uncertainties, while the statistical uncertainties are negligible.} 
    \label{fig:fig2}
\end{figure}

Figure~\ref{fig:fig2} (left) shows the charged-hadron \dnchdeta at midrapidity as a function of centrality.
The value of \dnchdeta at midrapidity ($\abs{\eta}<0.5$) is found to be $135.0 \pm 4.0\syst$ for events in the 0--5\% centrality class.
Previous measurements of \XeXe and \PbPb collisions at similar collision energies by the CMS~\cite{CMS:2024ykx,CMS:2019gzk} and ALICE~\cite{ALICE:2015juo,ALICE:2025cjn,ALICE:2018cpu} Collaborations are also shown.
The \OO data show a lower charged-hadron \dnchdeta compared to larger systems, as expected.
In the most central events, \dnchdeta of \OO collisions is about a factor of 15 smaller than in \PbPb collisions at $\sqrtsNN = 5.36\TeV$.
Moreover, the centrality dependence in \OO collisions is less steep than in \XeXe and \PbPb systems. 

Figure~\ref{fig:fig2} (right) presents the particle density normalized by $2A$, $(\dnchdeta)/2A$, as a function of centrality for the same data sets as in Fig.~\ref{fig:fig2} (left), where $A$ denotes the nuclear mass number.
Previous measurements have indicated that \dnchdeta approximately scales with $2A$ across ion species at fixed collision energy~\cite{CMS:2019gzk}.
The \OO results broadly follow this system-size trend but display clear departures from exact $2A$ scaling over centrality.  
This behavior reflects the system-dependent relation between centrality percentile and geometric overlap, whereby peripheral \OO collisions still correspond to comparatively large participant fractions.
For example, the 50--55\% centrality interval corresponds to $\avenpart/2A \approx 0.25$ in \OO collisions, compared to approximately 0.15 in \PbPb collisions.

\begin{figure}[hbt!]
  \centering
    \includegraphics[width=0.49\textwidth]{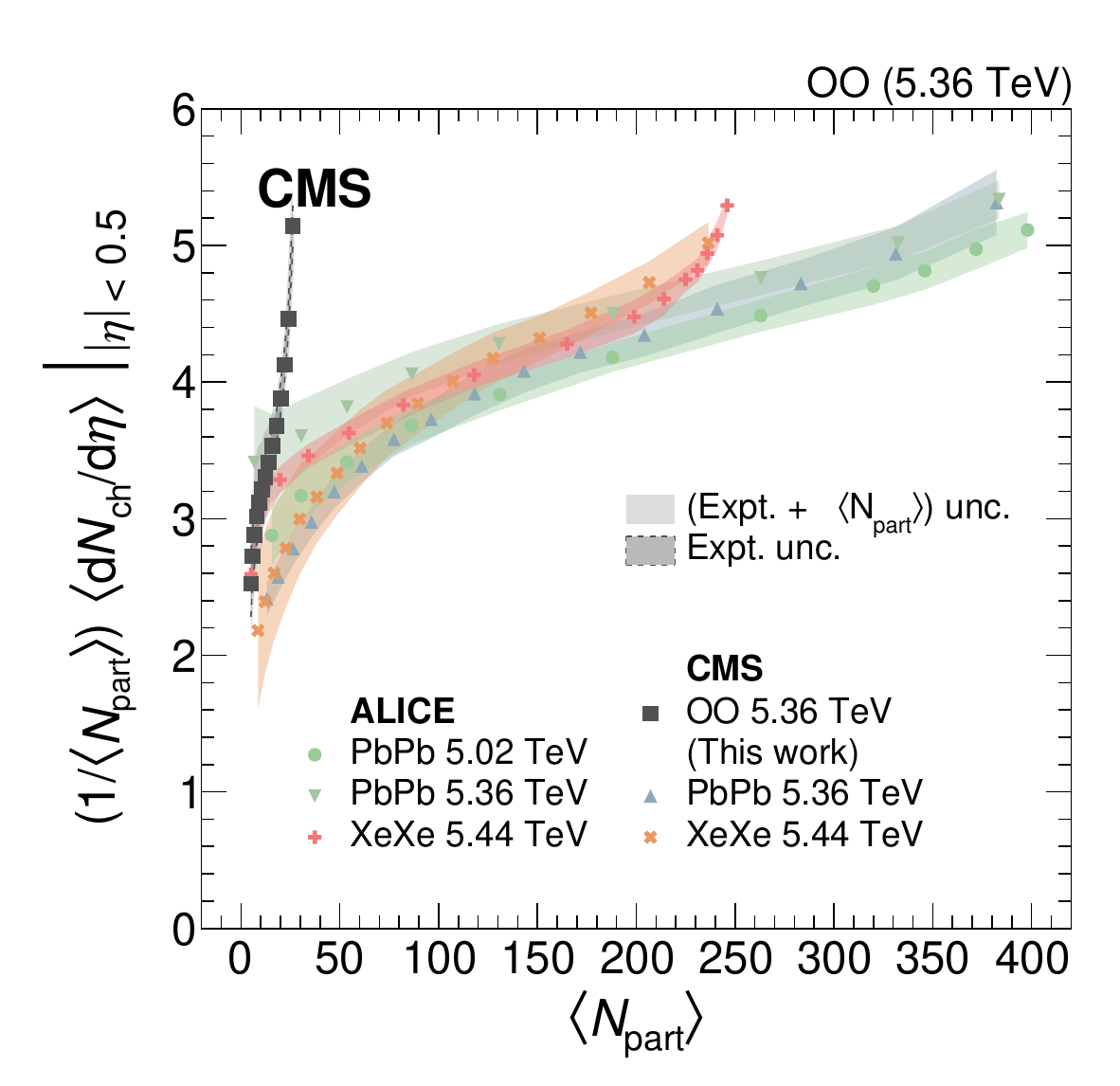}
    \includegraphics[width=0.49\textwidth]{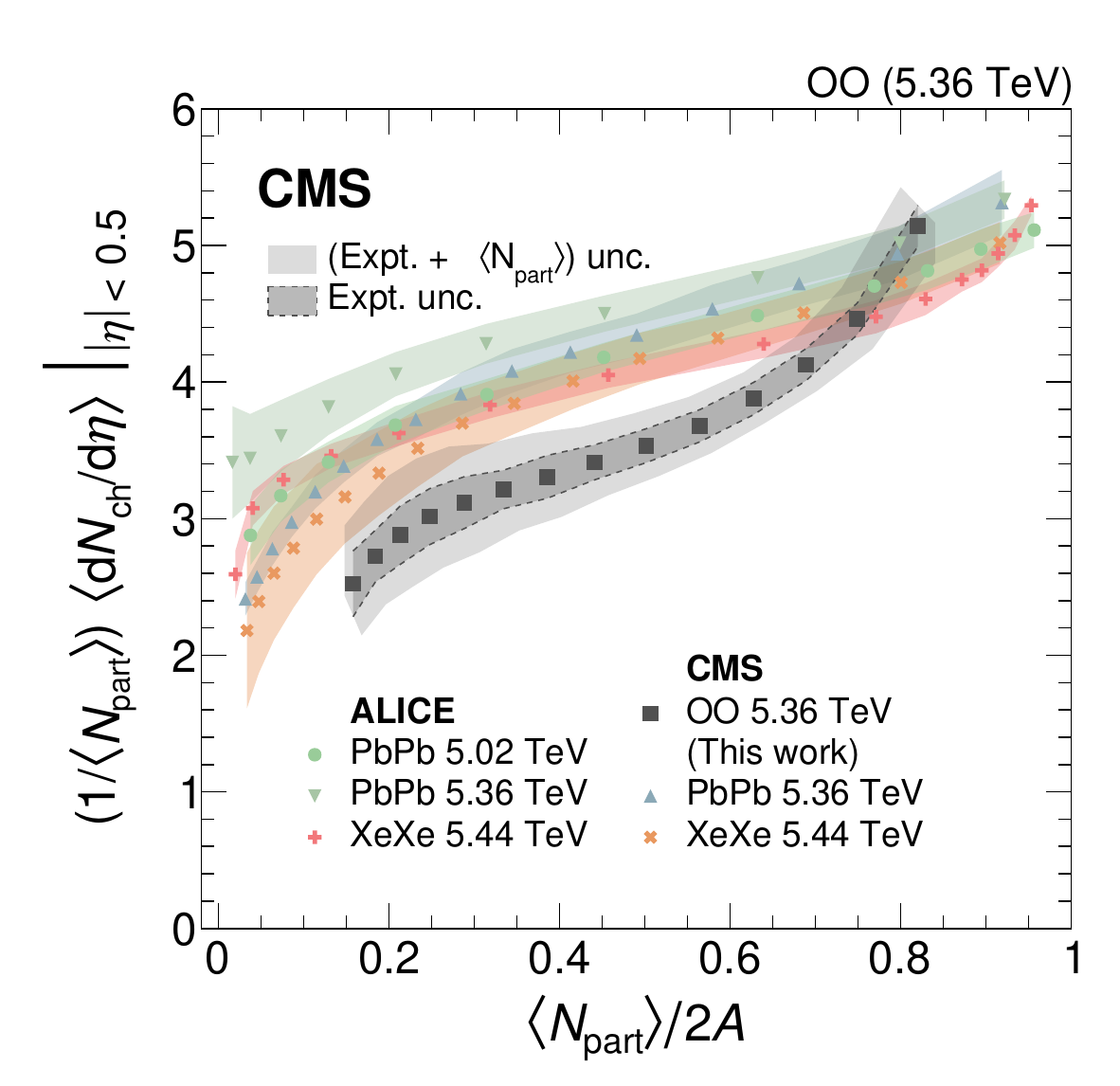}
    \caption{
    Charged-hadron \dnchdeta at midrapidity, normalized by \avenpart, in \OO collisions at $\sqrtsNN = 5.36\TeV$ shown as a function of \avenpart (\cmsLeft) and $\avenpart/2A$ (\cmsRight), where $A$ is the mass number of the colliding nuclei. 
    The results are compared with measurements in \PbPb and \XeXe collisions by the CMS~\cite{CMS:2024ykx,CMS:2019gzk} and ALICE~\cite{ALICE:2025cjn,ALICE:2015juo,ALICE:2018cpu} Collaborations.
    The bands around the data points denote the systematic uncertainties, while the statistical uncertainties are negligible.
    For the \OO data, the inner band shows the experimental uncertainties, while the outer band includes both the experimental and $\langle N_\mathrm{part}\rangle$ uncertainties.
    }
    \label{fig:fig3}
\end{figure}

The quantity $(\dnchdeta)/\avenpart$, representing the charged-hadron multiplicity per participating nucleon, is used to study particle production as a function of collision geometry.
It is shown as a function of \avenpart in Fig.~\ref{fig:fig3} (\cmsLeft) and as a function of $\avenpart/2A$, to account for the system size, in Fig.~\ref{fig:fig3} (\cmsRight).
As before, the \OO data are compared to measurements in \PbPb and \XeXe collisions by the CMS~\cite{CMS:2024ykx,CMS:2019gzk} and ALICE~\cite{ALICE:2025cjn,ALICE:2015juo,ALICE:2018cpu} Collaborations.
It is important to note that the same number of participants can correspond to different centrality classes across different collision systems. 
At similar collision energy, the distributions differ among \OO, \XeXe, and \PbPb collisions. 
The multiplicity per participant at a given \avenpart is significantly higher in central \OO collisions than in peripheral \XeXe and \PbPb collisions. 
A similar trend is observed when comparing central \XeXe to midcentral \PbPb collisions.
At a given $\avenpart/2A$, the $(\dnchdeta)/\avenpart$ for \OO is below that for both \XeXe and \PbPb. 
The small differences previously observed between \PbPb and \XeXe collisions, although compatible within uncertainties, suggested the importance of initial geometry for particle production at LHC energies~\cite{PHOBOS:2007hfc}. 
The \OO data show a stronger deviation from this behavior, both in the magnitude of the per-participant yields and in their dependence on $\avenpart/2A$.

\begin{figure}[hbt!]
  \centering
    \includegraphics[width=0.49\textwidth]{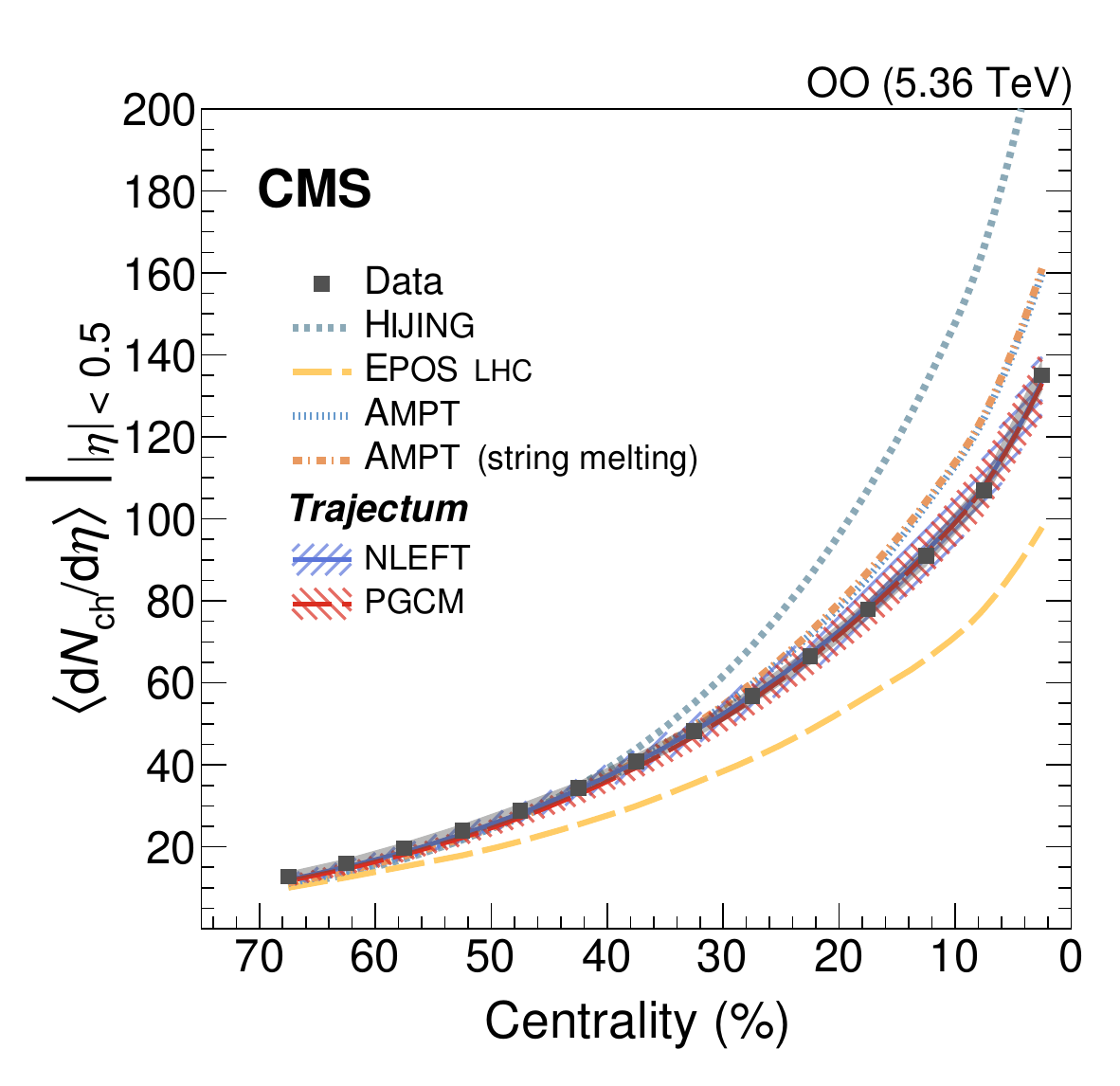}
    \includegraphics[width=0.49\textwidth]{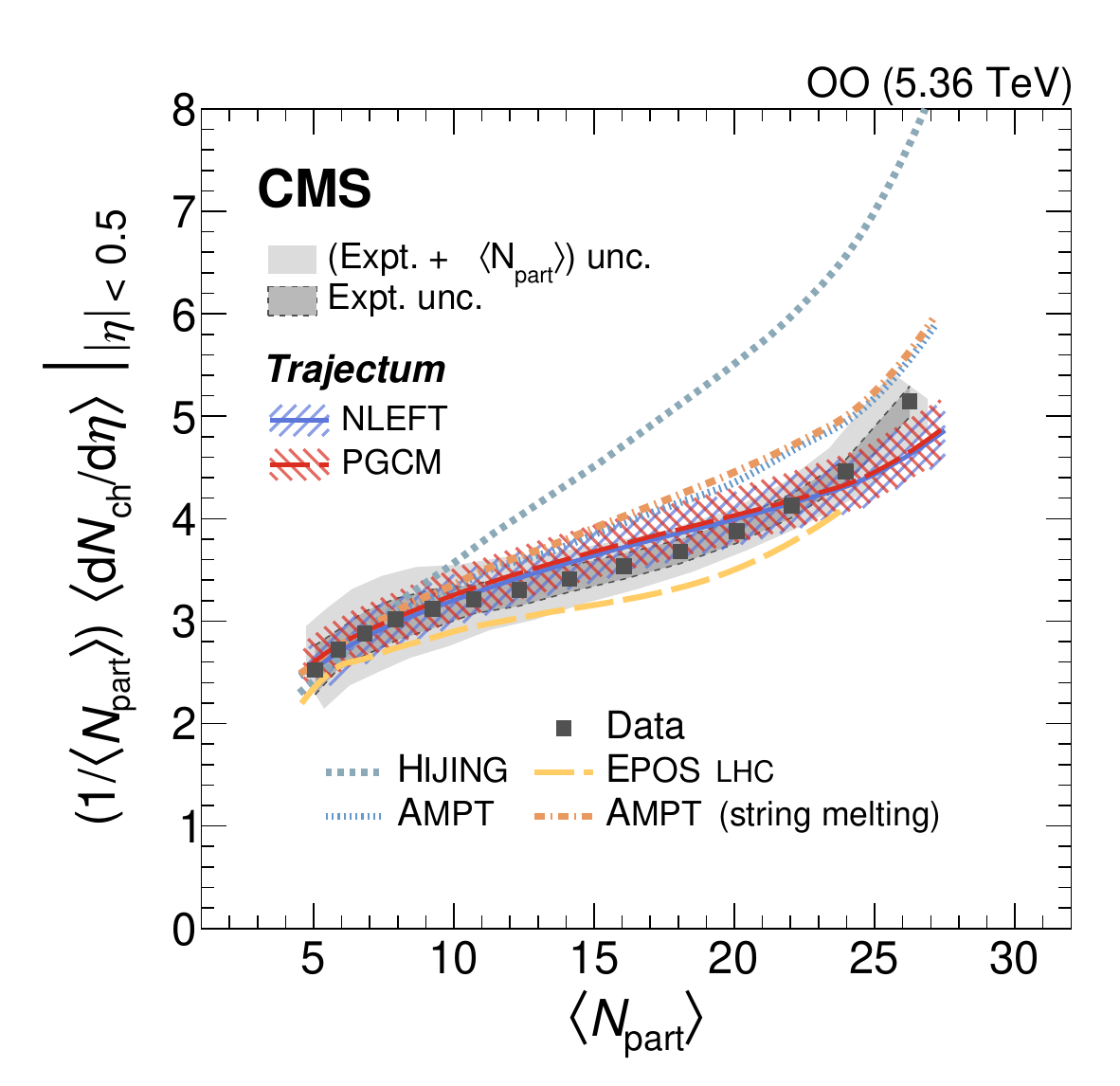}
    \caption{
    Charged-hadron \dnchdeta at midrapidity in \OO collisions at $\sqrtsNN = 5.36\TeV$ as a function of event centrality (\cmsLeft) and \dnchdeta normalized to \avenpart as a function of \avenpart (\cmsRight). 
    The results are compared to predictions from the \HIJING~\cite{Wang:1991hta}, \eposlhc~\cite{Werner:2005jf,Pierog:2013ria}, and \ampt~\cite{Lin:2004en} event generators, as well as the hydrodynamic model \trajectum~\cite{Giacalone:2024luz} using two different nuclear density descriptions for oxygen.
    The bands around the data points denote the systematic uncertainties, while the statistical uncertainties are negligible.
    For the \OO data, the inner band shows the experimental uncertainties, while the outer band includes both the experimental and $\langle N_\mathrm{part}\rangle$ uncertainties.
    }
    \label{fig:fig4}
\end{figure}

Figure~\ref{fig:fig4} compares the centrality-dependent \dnchdeta data at midrapidity for \OO collisions with predictions from the \HIJING~\cite{Wang:1991hta}, \eposlhc~\cite{Werner:2005jf,Pierog:2013ria}, and \ampt~\cite{Lin:2004en} event generators, as well as from the hydrodynamic model \trajectum~\cite{Giacalone:2024luz} using two different nuclear density descriptions for the oxygen nucleus: one obtained from nuclear lattice effective field theory (NLEFT) and another derived from the ab initio projected generator coordinate method (PGCM).
For the event generators, the centrality determination of simulated events is calibrated using $E_\mathrm{T}$ in the HF calorimeters, ensuring a direct comparison with the data. 
For the \trajectum calculations, however, event centrality is defined using the multiplicity of charged particles with $\pt > 0.4\GeV$ and $\abs{\eta} < 2.4$, as the model is primarily constrained by midrapidity observables rather than forward-energy distributions.

The left panel of Fig.~\ref{fig:fig4} displays \dnchdeta as a function of centrality, compared to the various model calculations. 
The \ampt generator is in reasonable agreement with the data, with little difference between the versions with and without string melting, while \eposlhc systematically underpredicts and \HIJING overpredicts the results in more central collisions.
The \trajectum model provides the best overall description of the data, predicting the results within about 10\% in peripheral and 1\% in central collisions. 
The NLEFT and PGCM predictions are nearly identical for the observables considered here.

The right panel of Fig.~\ref{fig:fig4} shows normalized $(\dnchdeta)/\avenpart$ data as a function of \avenpart. 
For the data, the \avenpart values are taken from Table~\ref{tab:npart}, 
whereas the models are plotted using the number of participating nucleons obtained in each simulation.
The \ampt model shows reasonable agreement with the data, while \HIJING overpredicts it, similar to the trend seen for the unnormalized \dnchdeta. 
The \trajectum calculations reproduce the normalized data well. 
The small deviation observed in the most central collisions is largely driven by differences in the corresponding \npart values, which may originate from differences in the underlying Glauber implementation or from differences in the centrality determination based on multiplicity.
In the case of \eposlhc, the agreement with the data improves significantly when comparing $(\dnchdeta)/\avenpart$, suggesting that while \eposlhc generates a realistic \dnchdeta distribution for a given initial condition, its description of the forward energy used for centrality determination differs from that observed in data.

\begin{figure}[htb!]
  \centering
     \includegraphics[width=\cmsFigWidth]{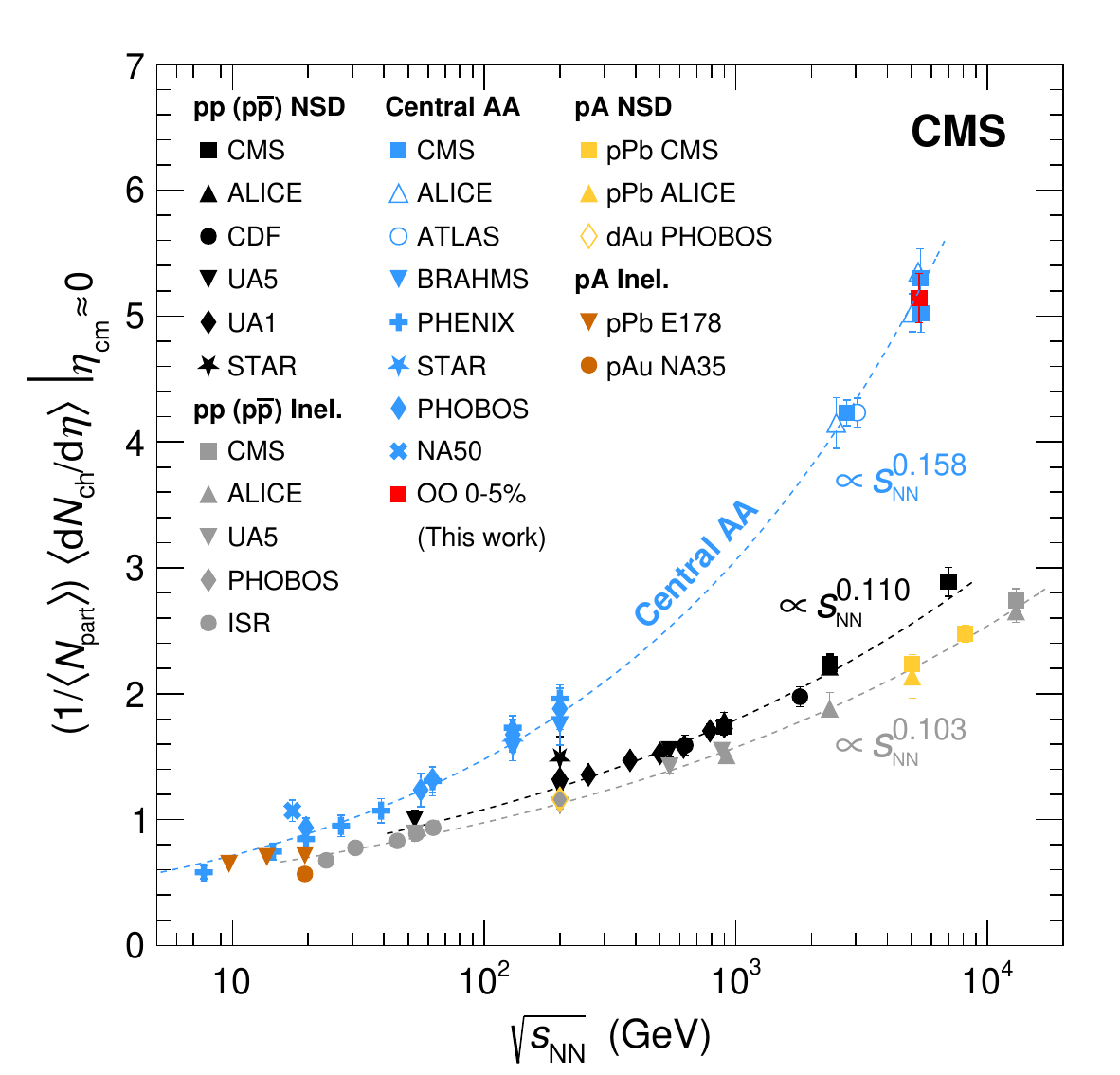}
    \caption{
    Comparison of average \dnchdeta at midrapidity, scaled by \avenpart in \pPb~\cite{ALICE:2012xs,CMS:2017shj}, \pAu~\cite{NA35:1997dow}, \dAu~\cite{PHOBOS:2003fjw,PHENIX:2015tbb,PHENIX:2017nae}, and central heavy ion collisions~\cite{CMS:2011aqh,CMS:2024ykx,CMS:2019gzk,PHOBOS:2010eyu,NA50:2002edr,STAR:2001eyo,BRAHMS:2001gci,BRAHMS:2001llo,PHENIX:2000owy,PHOBOS:2000wxz,PHOBOS:2001zjw,ALICE:2010khr,ATLAS:2011ag,ALICE:2025cjn,PHENIX:2015tbb,STAR:2009sxc,STAR:2008med}, as well as nonsingle-diffractive (NSD)~\cite{CMS:2010wcx,CMS:2010tjh,STAR:2008med,UA1:1989bou,UA5:1982yte,CDF:1989nkn} and inelastic~\cite{CMS:2015zrm,PHOBOS:2010eyu,Aachen-CERN-Heidelberg-Munich:1977izz,ALICE:2010cin} \pp collisions. 
    The data points for nucleus-nucleus (\AandA) collisions at $\sqrtsNN = 2.76\TeV$ have been shifted horizontally by $\pm$10\% for visibility. 
    The dashed curves, reproduced from Ref.~\cite{CMS:2017shj}, are included to guide the eye and correspond to a power law functional form.}
    \label{fig:fig5}
\end{figure}

Figure~\ref{fig:fig5} further explores particle production by examining the dependence of \dnchdeta normalized by \avenpart on the collision energy for various collision systems and event selections. 
Power law fits to previous measurements reproduced from Ref.~\cite{CMS:2017shj} are shown as dashed lines to guide the eye.  
It has been previously observed that central \AandA collisions yield larger values than nonsingle-diffractive (NSD) \pp collisions, indicating a more efficient conversion of the initial collision energy into final-state particle multiplicity in nucleus-nucleus systems. 
The new CMS data from \OO collisions, shown by the red point at $\sqrtsNN=5.36\TeV$, are consistent with this scaling behavior observed at lower energies. 
This result indicates that particle production in central \OO collisions follows the same enhanced scaling seen in heavier ion systems, supporting the presence of collective dynamics beyond independent nucleon-nucleon interactions.
 
\section{Summary}
The pseudorapidity ($\eta$) distributions of charged hadrons are measured for the first time in oxygen-oxygen (\OO) collisions at a center-of-mass collision energy per nucleon pair $\sqrtsNN = 5.36\TeV$.  
The measurements are performed in the range $\abs{\eta}<2.4$ for multiple centrality intervals. 
The charged-hadron pseudorapidity density in the midrapidity region ($\abs{\eta}<0.5$) is measured to be $\langle \rd N_{\text{ch}}/\rd \eta \rangle = 41.8 \pm 1.1\syst$ integrated over centrality.
The $\eta$ dependence of the centrality-integrated charged-hadron pseudorapidity density is compared to predictions from the \HIJING, \eposlhc, and \ampt event generators. 
Among these calculations, \ampt best reproduces the absolute yields, while \eposlhc provides the best description of the $\eta$ dependence.  

For the 5\% most central collisions, the charged-hadron density at midrapidity ($\abs{\eta} < 0.5$) is measured to be $\dnchdeta = 135.0 \pm 4.0\syst$. 
The corresponding per-participant charged-hadron density is consistent with that observed in central \PbPb collisions at similar collision energies, indicating that particle production per participating nucleon in central collisions remains comparable even in substantially smaller collision systems.

The centrality dependence of the midrapidity results is compared with previous measurements in lead-lead (\PbPb) collisions at $\sqrtsNN = 5.02$ and 5.36\TeV and xenon-xenon (\XeXe) collisions at $\sqrtsNN = 5.44$\TeV. 
The expected system-size ordering is observed, with \dnchdeta increasing from \OO to \XeXe to \PbPb at fixed centrality. 
After normalization by the maximum number of nucleons in the colliding system, the \OO results follow this trend approximately, but exhibit a less steep centrality dependence.
At a given participant fraction, the per-participant yields are consistent between \PbPb and \XeXe collisions, while those in \OO collisions show a clear deviation.
These observations indicate that, while the overall energy-scaling behavior observed in nucleus-nucleus collisions is preserved, the centrality dependence of normalized particle production deviates from the trends observed in larger collision systems. 
This behavior highlights the increased importance of collision geometry and finite-size effects in the smallest ion-ion system studied at the LHC. 
The hydrodynamic model \trajectum provides the best overall description of the midrapidity data, particularly in central collisions. 
The \OO data provide new constraints on models of initial-state entropy production and particle production dynamics in light ion-ion collisions.

\begin{acknowledgments}
We congratulate our colleagues in the CERN accelerator departments for the excellent performance of the LHC and thank the technical and administrative staffs at CERN and at other CMS institutes for their contributions to the success of the CMS effort. In addition, we gratefully acknowledge the computing centres and personnel of the Worldwide LHC Computing Grid and other centres for delivering so effectively the computing infrastructure essential to our analyses. Finally, we acknowledge the enduring support for the construction and operation of the LHC, the CMS detector, and the supporting computing infrastructure provided by the following funding agencies: SC (Armenia), BMBWF and FWF (Austria); FNRS and FWO (Belgium); CNPq, CAPES, FAPERJ, FAPERGS, and FAPESP (Brazil); MES and BNSF (Bulgaria); CERN; CAS, MoST, and NSFC (China); MINCIENCIAS (Colombia); MSES and CSF (Croatia); RIF (Cyprus); SENESCYT (Ecuador); ERC PRG and PSG, TARISTU24-TK10 and MoER TK202 (Estonia); Academy of Finland, MEC, and HIP (Finland); CEA and CNRS/IN2P3 (France); SRNSF (Georgia); BMFTR, DFG, and HGF (Germany); GSRI (Greece); MATE and NKFIH (Hungary); DAE and DST (India); IPM (Iran); SFI (Ireland); INFN (Italy); MSIT and NRF (Republic of Korea); MES (Latvia); LMTLT (Lithuania); MOE and UM (Malaysia); BUAP, CINVESTAV, CONACYT, LNS, SEP, and UASLP-FAI (Mexico); MOS (Montenegro); MBIE (New Zealand); PAEC (Pakistan); MSHE, NSC, and NAWA (Poland); FCT (Portugal); MESTD (Serbia); MICIU/AEI and PCTI (Spain); MOSTR (Sri Lanka); Swiss Funding Agencies (Switzerland); MST (Taipei); MHESI (Thailand); TUBITAK and TENMAK (T\"{u}rkiye); NASU (Ukraine); STFC (United Kingdom); DOE and NSF (USA).

\hyphenation{Rachada-pisek} Individuals have received support from the Marie-Curie programme and the European Research Council and Horizon 2020 Grant, contract Nos.\ 675440, 724704, 752730, 758316, 765710, 824093, 101115353, 101002207, 101001205, and COST Action CA16108 (European Union); the Leventis Foundation; the Alfred P.\ Sloan Foundation; the Alexander von Humboldt Foundation; the Science Committee, project no. 22rl-037 (Armenia); the Fonds pour la Formation \`a la Recherche dans l'Industrie et dans l'Agriculture (FRIA) and Fonds voor Wetenschappelijk Onderzoek contract No. 1228724N (Belgium); the Beijing Municipal Science \& Technology Commission, No. Z191100007219010, the Fundamental Research Funds for the Central Universities, the Ministry of Science and Technology of China under Grant No. 2023YFA1605804, the Natural Science Foundation of China under Grant No. 12535004, and USTC Research Funds of the Double First-Class Initiative No.\ YD2030002017 (China); the Ministry of Education, Youth and Sports (MEYS) of the Czech Republic; the Shota Rustaveli National Science Foundation (Georgia); the Deutsche Forschungsgemeinschaft (DFG), among others, under Germany's Excellence Strategy -- EXC 2121 ``Quantum Universe" -- 390833306, and under project number 400140256 - GRK2497; the Hellenic Foundation for Research and Innovation (HFRI), Project Number 2288 (Greece); the Hungarian Academy of Sciences, the New National Excellence Program - \'UNKP, the NKFIH research grants K 131991, K 138136, K 143460, K 143477, K 147557, K 146913, K 146914, K 147048, TKP2021-NKTA-64, and 2025-1.1.5-NEMZ\_KI-2025-00004, and MATE KKP and KKPCs Research Excellence and Flagship Research Groups grants (Hungary); the Council of Science and Industrial Research, India; ICSC -- National Research Centre for High Performance Computing, Big Data and Quantum Computing, FAIR -- Future Artificial Intelligence Research, and CUP I53D23001070006 (Mission 4 Component 1), funded by the NextGenerationEU program, the Italian Ministry of University and Research (MUR) under Bando PRIN 2022 -- CUP I53C24002390006, PRIN PRIMULA 2022RBYK7T (Italy); the Latvian Council of Science; the Ministry of Science and Higher Education, project no. 2022/WK/14, and the National Science Centre, contracts Opus 2021/41/B/ST2/01369, 2021/43/B/ST2/01552, 2023/49/B/ST2/03273, and the NAWA contract BPN/PPO/2021/1/00011 (Poland); the Funda\c{c}\~ao para a Ci\^encia e a Tecnologia (Portugal); the National Priorities Research Program by Qatar National Research Fund;  MICIU/AEI/10.13039/501100011033, ERDF/EU, ``European Union NextGenerationEU/PRTR", projects PID2022-142604OB-C21, PID2022-139519OB-C21, PID2023-147706NB-I00, PID2023-148896NB-I00, PID2023-146983NB-I00, PID2023-147115NB-I00, PID2023-148418NB-C41, PID2023-148418NB-C42, PID2023-148418NB-C43, PID2023-148418NB-C44, PID2024-158190NB-C22, RYC2021-033305-I, RYC2024-048719-I, CNS2023-144781, CNS2024-154769 and Plan de Ciencia, Tecnolog{\'i}a e Innovaci{\'o}n de Asturias, Spain; the Chulalongkorn Academic into Its 2nd Century Project Advancement Project, the National Science, Research and Innovation Fund program IND\_FF\_68\_369\_2300\_097, and the Program Management Unit for Human Resources \& Institutional Development, Research and Innovation, grant B39G680009 (Thailand); the Eric \& Wendy Schmidt Fund for Strategic Innovation through the CERN Next Generation Triggers project under grant agreement number SIF-2023-004; the Kavli Foundation; the Nvidia Corporation; the SuperMicro Corporation; the Welch Foundation, contract C-1845; and the Weston Havens Foundation (USA).
\end{acknowledgments}\section*{Data availability} Release and preservation of data used by the CMS Collaboration as the basis for publications is guided by the  \href{https://doi.org/10.7483/OPENDATA.CMS.1BNU.8V1W}{CMS data preservation, re-use and open access policy}.
\bibliography{auto_generated}
\cleardoublepage \appendix\section{The CMS Collaboration \label{app:collab}}\begin{sloppypar}\hyphenpenalty=5000\widowpenalty=500\clubpenalty=5000\input{HIN-25-010-public-authorlist.tex}\end{sloppypar}
\end{document}

%% file: HIN-25-010-public-authorlist.tex
\cmsinstitute{Yerevan Physics Institute, Yerevan, Armenia}
{\tolerance=6000
A.~Belyaev\cmsorcid{0000-0003-1692-1173}, A.~Gevorgyan\cmsorcid{0000-0003-2751-9489}, A.~Hayrapetyan, A.~Tumasyan\cmsAuthorMark{1}\cmsorcid{0009-0000-0684-6742}
\par}
\cmsinstitute{Marietta Blau Institute for Particle Physics, Vienna, Austria}
{\tolerance=6000
W.~Adam\cmsorcid{0000-0001-9099-4341}, L.~Benato\cmsorcid{0000-0001-5135-7489}, T.~Bergauer\cmsorcid{0000-0002-5786-0293}, M.~Dragicevic\cmsorcid{0000-0003-1967-6783}, S.~Gundacker\cmsorcid{0000-0003-2087-3266}, A.K.~Guven\cmsorcid{0009-0004-5670-5138}, P.S.~Hussain\cmsorcid{0000-0002-4825-5278}, M.~Jeitler\cmsAuthorMark{2}\cmsorcid{0000-0002-5141-9560}, N.~Krammer\cmsorcid{0000-0002-0548-0985}, A.~Li\cmsorcid{0000-0002-4547-116X}, D.~Liko\cmsorcid{0000-0002-3380-473X}, M.~Matthewman, A.~Pfeiffer\cmsorcid{0000-0001-5328-448X}, J.~Schieck\cmsAuthorMark{2}\cmsorcid{0000-0002-1058-8093}, R.~Sch\"{o}fbeck\cmsAuthorMark{2}\cmsorcid{0000-0002-2332-8784}, M.~Shooshtari\cmsorcid{0009-0004-8882-4887}, M.~Sonawane\cmsorcid{0000-0003-0510-7010}, N.~Van~Den~Bossche\cmsorcid{0000-0003-2973-4991}, W.~Waltenberger\cmsorcid{0000-0002-6215-7228}, C.E.~Wulz\cmsAuthorMark{2}\cmsorcid{0000-0001-9226-5812}
% ...
\par}
$^{1}$Also at Yerevan State University, Yerevan, Armenia\\
$^{2}$Also at Technische Universit\"{a}t Wien, Vienna, Austria\\
% ...
